\documentstyle[preprint,aps,epsf]{revtex}

\tighten
\begin{document}
\bibliographystyle{/usr/share/texmf/tex/latex/revtex/prsty}
\draft
\title{ 
Benchmark Test Calculation  of  a  Four-Nucleon Bound  State  
}

\author{H. Kamada
\footnote{present address: Forschungszentrum J\"ulich, Institut f\"ur Kernphysik (Theorie), 
D-52425 J\"ulich, Germany,
email:kamada@tp2.ruhr-uni-bochum.de},
A. Nogga and W. Gl\"ockle
}
\address{
Institut f\"ur  Theoretische Physik II,
         Ruhr-Universit\"at Bochum, 44780 Bochum, Germany}

\author{E. Hiyama$^{\S}$ and M. Kamimura$^{\S\S}$}

\address{$^{\S}$ High Energy Accelerator Research Organization, Institute of Particle and
                            Nuclear Studies, Japan }
\address{$^{\S\S}$ Department of Physics, Kyushu University, Fukuoka 812-8581, Japan }

\author{K. Varga$^{**}$ and Y. Suzuki$^{***}$ }

\address{$^{**}$Solid State Division, Oak Ridge National Laboratory,
Oak Ridge, TN 37380}
\address{$^{**}$
Institute of Nuclear Research of the Hungarian Academy
of Sciences (ATOMKI) , Debrecen, 4000,  PO Box 51. Hungary}

\address{$^{***}$Department of Physics, Niigata University, Niigata 950-2181, Japan }

\author{M. Viviani$^{\sharp}$, A. Kievsky$^{ \sharp}$, and  S. Rosati$^{\sharp , \flat}$ }

\address{$^{\sharp}$INFN, Sezione di Pisa, I-56100 Pisa, Italy  }
\address{$^{\flat}$ Department of Physics, University of Pisa, I-56100 Pisa, Italy}

\author{J. Carlson$^{\%}$, Steven C. Pieper$^{\%\%}$ and R. B. Wiringa$^{\%\%}$}
\address{$^{\%}$Theoretical Division, Los Alamos National Laboratory, Los Alamos, New Mexico 87545 }
\address{$^{\%\%}$Physics Division, Argonne National Laboratory, Argonne, Illinois 60439 }

\author{P. Navr\'atil$^{a}$ and  B. R. Barrett$^{+}$}
\address{$^{a}$ Lawrence Livermore National Laboratory, P. O. Box 808, Livermore, CA 94551} 
\address{$^{a}$ Nuclear Physics Institute, 
Academy of Sciences of the Czech Republic, 250 68 \v{R}e\v{z} near Prague, Czech Republic}
\address{$^{+}$ Department of Physics, P.O. Box 210081, University of Arizona, Tucson, AZ, 85721 USA  }

\author{N. Barnea$^{++}$, W. Leidemann$^{+++}$ and G. Orlandini$^{+++}$}

\address{$^{++}$The Racah Institute of Physics, The Hebrew University, 91904,
Jerusalem, Israel}
\address{$^{+++}$Dipartimento di Fisica and INFN (Gruppo Collegato di Trento), 
Universit\`a di Trento, I-38050 Povo, Italy }

\date{\today}

\maketitle

\begin{abstract}
In the past, several 
efficient methods have been developed 
to solve the Schr\"odinger equation for four-nucleon bound  states 
accurately. These are the  Faddeev-Yakubovsky, the coupled-rearrangement-channel 
Gaussian-basis variational, the stochastic variational, 
the hyperspherical variational,  the Green's function Monte  Carlo,
the no-core shell model and the effective interaction hyperspherical harmonic
methods.
In this article we compare the 
energy eigenvalue results and some wave function properties using 
the realistic AV8$'$ NN interaction.
The results of all schemes agree very well
showing the high accuracy of our present ability to calculate  the four-nucleon
bound state.
\end{abstract}


\narrowtext

\section{Introduction}

Recent advances in computational facilities,
together with the development of new methods and refinements upon 
older ones, allow  very precise  
calculations for few-body systems. 
These advances are especially remarkable 
in nuclear physics considering the complexity of the nuclear interaction. 
The three-nucleon(3N)  bound-state \cite{Nogga97,PISA1,Krivec}
and scattering-state\cite{Gloeckle,PISA2,CHEN1} problems  
are rigorously solved using realistic nuclear  
potentials\cite{cdBonn,AV18,Nijm}.
These calculational schemes are mostly based on a partial wave 
decomposition. 
Stochastic and Monte Carlo methods for bound states, however,  
are performed directly
using position vectors in configuration  space. 
Also in momentum space the first steps have been taken to avoid partial wave 
decompositions in both two-nucleon (NN)~\cite{Fachruddin} and  
3N\cite{Elster,Schadow} systems. 
Benchmark calculations based on different algorithms 
for the 3N continuum both below\cite{Friar90,Kievsky98}
and above\cite{Friar95} the deuteron threshold already exist.

The complicated calculation of few-body continuum states can be avoided
in the evaluation of reaction cross sections, even in the presence of
realistic forces \cite{Efros00}. In fact the transition strength can be 
calculated in an alternative way, where only bound state techniques are 
needed \cite{Efros94}. 

There are a few analytical solutions of 3N  bound states\cite{Jensen} for 
square-well potentials,
against which numerical solutions have been checked, but they are 
far from possessing the complexity of realistic nuclear forces.
In the four-body system we are only aware of benchmark calculations for four bosons~\cite{Krivec}.

Benchmark calculations are extremely useful to test methods as well as
calculational schemes.  
They are also  often of interest for a general readership, since they may
help to solve analogous problems in other fields. 
We think that this is
particularly the case for the quite complex four-fermion system.
Here we would like to address the 
four-nucleon (4N) bound-state problem 
using the AV8$'$ NN potential\cite{AV18+} which is a simplified, reprojected version of the 
fully realistic Argonne AV18 model\cite{AV18}, 
but still has most of its complexity, e.g.,  the tensor force is built in.

In Sec. 2 the different methods are briefly introduced and the results
are presented in Sec. 3 together  with a brief summary.

\section{Methods}

In order to solve the bound four-nucleon  system we employ the   
Faddeev-Yakubovsky equations(FY)\cite{Yakubovsky,kamada,nogga00,Gloeckletext},
the Coupled-Rearrangement-Channel 
Gaussian-Basis Variational Method
(CRCGV)\cite{Kamimura88,Kameyama89,Kamimura90,Kino,Hiyama99,Hiyama00a,Hiyama00b,Kino96}, 
the Stochastic Variational Method(SVM) with correlated Gaussians\cite{c1,c2,c3,c4}, 
the Hyperspherical Harmonic Variational Method(HH)\cite{A89,F83,VKR95,KMRV96,V98,KRV01},
the Green's function Monte Carlo(GFMC)\cite{Carlson,Carlson2,AV18+,aequal8} 
method, 
the No-Core 
Shell Model(NCSM)\cite{Navratil,LS80,LSeffop}, and the effective interaction hyperspherical 
harmonic method(EIHH)\cite{Barnea}.
The various procedures are briefly described below.

\subsection{Faddeev-Yakubovsky Equations}

The 4N Faddeev-Yakubovsky equations in momentum space are                
\cite{kamada,nogga00,Gloeckletext}  
\begin{eqnarray}
\psi_1 = G_0 t_{12} P [ (1- P_{34})\psi_1 + \psi_2 ],
\\
\psi_2= G_0 t_{12} \tilde P [ (1 - P_{34}) \psi_1 + \psi_2 ],
\end{eqnarray}
where $\psi_1$ and $\psi_2$ are Yakubovsky components and $t_{12} $ is the 
two nucleon transition 
matrix determined by a  two nucleon Lippmann - Schwinger equation.
 $P$, $P_{34}$ and $\tilde P$ 
are permutation operators: $ P= P_{12}P_{23} + P_{13} P_{23}$, 
$\tilde P= P_{13}P_{24} $,  
where the $P_{ij}$ are transpositions of particles $i$ and $j$. 
The fully anti-symmetrized wave function $\Psi$  is  
\begin{eqnarray}
\Psi = [1 - (1+P) P_{34} ] (1 + P) \psi_1 + (1+P)(1+\tilde P)\psi_2 .
\end{eqnarray}
The Yakubovsky equations are decomposed into partial waves. 
We truncate the partial waves at 
a 2-body total angular momentum $j \le 6 $, 
all other orbital angular momenta at $l_i \le 8$ ,  and the sum of all angular 
momenta at $\sum_i l_i \le 12$. In this truncation we keep  1572 angular momentum and 
isospin combinations (often called channels).
This is sufficient to guarantee convergence of our results as given in 
Sec. 3. 
The diagonalization is performed by  a modified Lanczos method\cite{saake}.
Recent results for realistic NN potentials, and including three-nucleon forces,  
are given  in \cite{nogga00}.

\subsection{Coupled-Rearrangement-Channel 
Gaussian-Basis Variational Method}

The coupled-rearrangement-channel Gaussian basis variational
method was proposed by Kamimura 
\cite{Kamimura88} 
to solve the Coulombic  three-body problem 
of the muonic molecular ion $(dt\mu)^+$, 
within an accuracy of seven significant figures for the  
energy of the very loosely bound $J=v=1$ state; 
this accuracy was required for the comparison with 
experimental data on the muon catalyzed fusion cycle.
Use of basis functions that spanned all the  three
rearrangement Jacobian coordinates was essential to the 
high-precision calculation.
The method was also applied to three-nucleon bound states
\cite{Kameyama89,Kamimura90}  
and was found to accomplish a much more rapid convergence 
in the binding energy 
with respect to the number of the three-body 
angular momentum channels 
(see Fig. 5 of
\cite{Kamimura90}). 

The method was also successfully used to make another 
high-precision
Coulomb three-body calculation of the  
anti-protonic helium atom ($\bar{p}+{\rm He}^{++}+e^-$)  
in highly excited meta-stable 
states with $J\approx35$
\cite{Kino}. 
The calculation agreed with the high-resolution 
laser spectroscopic data 
within seven significant figures so that 
the mass of antiproton was 
derived to two orders of magnitude better precision
than published values.  
The method has been useful in four-body calculations of
the structure of light hypernuclei with realistic $YN$ and $NN$
interactions \cite{Hiyama99,Hiyama00a,Hiyama00b}. 

The total four-body wave function is described as the sum of amplitudes of
the rearrangement-Jacobian-coordinate channels with the $LS$
coupling scheme : 
\begin{eqnarray}
 \Psi_{JM} =  
\sum_{\alpha}  C^{(K)}_\alpha \Phi^{(K)}_\alpha +  
\sum_{\alpha}  C^{(H)}_\alpha \Phi^{(H)}_\alpha \; ,
\end{eqnarray}
where anti-symmetrized  basis functions
are described with  quantum numbers
$\alpha \equiv \{nl, NL, \Lambda, \nu \lambda, I, s s'S, t \}$ 
 by 
\begin{eqnarray}
&&\Phi^{(K)}_\alpha =        {\cal A}
 \left\{ 
\Big[ \big[ [\phi_{nl}({\bf r}) 
         \psi_{NL}({\bf R})]_\Lambda \;
        \varphi_{\nu\lambda} 
         (\mbox{\boldmath $\rho$}) \big]_{I}        
     \right. \cr    
        &\times&
          \big[ [\chi_s(12)
        \chi_{\frac{1}{2}}(3)]_{s'} 
        \chi_{\frac{1}{2}}(4) 
          \big]_{S}    \Big]_{JM} 
    \left.     
          \big[ [\eta_t(12)
        \eta_{\frac{1}{2}}(3)]_{\frac{1}{2}} 
        \eta_{\frac{1}{2}}(4) 
          \big]_0 \right\} ,
\end{eqnarray}
\begin{eqnarray}
\Phi^{(H)}_\alpha =        {\cal A}
 \left\{ 
\Big[  \big[[\phi_{nl}({\bf r}') 
         \psi_{NL}({\bf R}')]_\Lambda \;
        \varphi_{\nu\lambda} 
         (\mbox{\boldmath $\rho$}') \big]_{I}        
     \right.         \nonumber
\end{eqnarray}
\begin{eqnarray}
        \times 
          \big[ \chi_s(12)\chi_{s'}(34)
          \big]_{S}    \Big]_{JM} 
    \left.     
          \big[ \eta_t(12)\eta_t(34)
          \big]_0 \right\} .   
\end{eqnarray}
We employ  $K$-type coordinates 
${\bf r}= {\bf x}_1-{\bf x}_2, 
{\bf R}=({\bf x}_1+{\bf x}_2)/2-{\bf x}_3$,
$\mbox{\boldmath $\rho$}=({\bf x}_1+{\bf x}_2+{\bf x}_3)/3-{\bf x}_4$ 
and $H$-type ones 
${\bf r}'= {\bf x}_1-{\bf x}_2, {\bf R}'= {\bf x}_3-{\bf x}_4$,
$\mbox{\boldmath $\rho$}'=({\bf x}_1+{\bf x}_2)/2
-({\bf x}_3+{\bf x}_4)/2.$
$\cal{A}$ is the four-nucleon
antisymmetrizer and
$\chi$'s and $\eta$'s are the spin and isospin functions,
respectively. 
The functional form of 
$\phi_{nl}({\bf r})$ is taken as
\begin{eqnarray}
      \phi_{nlm}({\bf r})
      &=&
      r^l \, e^{-(r/r_n)^2} 
       Y_{lm}({\widehat {\bf r}})  \quad , 
\end{eqnarray}
where the Gaussian range parameters are chosen to lie in a
geometrical progression ($ r_n=
      r_1 a^{n-1}; n=1 \sim n_{\rm max}$),
and similarly for the other functions $\psi$ and
$\varphi$.
This manner of choosing  the range parameters 
is very suitable for describing 
both the short-range correlations and the long-range
asymptotic behavior precisely 
\cite{Kameyama89,Kino96}.

  Eigen-energies and wave-function  
coefficients $C$'s are determined by
solving the Schr\"{o}dinger equation with 
the Rayleigh-Ritz variational principle. 
It is to be emphasized that 
truncation  is not made for the partial waves of the 
$NN$ interaction, in contrast to  the Faddeev-Yakubovsky method, 
but is done only for the angular momenta of basis functions,
as in most variational methods.  This makes it possible to
accomplish a very quick convergence; the  result in Sec. 3. uses 
$l, L, \lambda \leq 2$
(this is the same as in the case of the three nucleon bound states,
mentioned above).  For instance, this amounts to 100 channels for the 
calculation in Sec. 3.

\subsection{Stochastic Variational Method}

The Correlated Gaussian trial function is written in the following form
\cite{c1,c2,c3}:
\begin{equation}
\Psi=\sum_{i=1}^{\cal K} c_i
{\cal A} \left\lbrace \left[\theta_{Li}({\hat {\bf x}})\xi_{Si}\right]_{JM}
\xi_{TM_Ti} {\rm exp}\left(-{1\over 2} {\bf x}A_i{\bf x}\right)
\right\rbrace,
\end{equation}
where ${\cal A}$ is the antisymmetrizer, ${\bf x}$ stands for a set of 
$A-1$ intrinsic coordinates $({\bf x}_1, {\bf x}_2,\ldots,{\bf x}_{A-1})$ 
and 
$\xi_{SMi}$ ($\xi_{TM_Ti}$) is the spin (isospin) function of 
the $A$-particle system. These functions are constructed by successively
coupling the spin (isospin) of the nucleons
\begin{equation}
\xi_{SMi}=[[[\chi_{1\over 2}(1)\chi_{1\over 2}(2)]_{s_{12}} 
\chi_{1\over 2}(3)]_{s_{123}}\ldots ]_{SMi}
\end{equation}
(similarly for the isospin part). 
The non-spherical (orbital) part of the trial function is represented by 
a successively coupled product of spherical harmonics 
\begin{equation}
\theta_{LMi}({\hat {\bf x}})=\left[\left[\left[
Y_{l_1}({\hat {\bf x}_1}) Y_{l_2}({\hat {\bf x}_2})\right]_{l_{12}} 
Y_{l_3}({\hat {\bf x}_3})\right]_{l_{123}}\ldots \right]_{LMi}.
\end{equation}
The index $i$ in the above equation stands for a label to distinguish 
the different possible intermediate coupling schemes as well as
the different possible total spin and total orbital angular momentum. 
\par\indent
The expansion over the partial waves has to be truncated, but the 
correlations included in the Gaussian part, ${\rm exp}(-{1\over 2} {\bf x}A_i{\bf x})$, make the trial function
flexible enough, so these truncations are not expected to seriously 
affect the accuracy. 
In the present calculation we included all partial
waves up to 
\begin{equation}
\sum_{i=1}^{A-1} l_i \le 4 .
\end{equation}
The trial function contains $A(A-1)/2$ nonlinear variational 
parameters. The total spin, total orbital angular momentum and
intermediate coupling quantum numbers are also variational parameters 
in the sense that one has to include all possibilities which improve 
the energy. We have a large number of parameters to be optimized and
it  is not at all clear how to select the optimal quantum numbers.
\par\indent
This variational basis is nonorthogonal, none of the components 
is indispensable, and one can replace a component by a linear combination 
of others. This gives us an excellent opportunity to use a stochastic
optimization procedure. To optimize the variational basis we used the
``stochastic variational method'' 
\cite{c1,c2,c3,c4}. In the SVM
one searches for the best wave function by a random trial and error
procedure. Random trial functions are generated and their energies are 
compared. Randomness in this case means that the quantum numbers 
and the nonlinear parameters are random numbers. Trial functions 
giving the lowest energy are selected as basis states. 
Details and various applications of the approach 
can be found in \cite{c1,c2,c3,c4}.
\par\indent
  The number of basis states used in the calculations is about 150 
for the triton and 300 for the  alpha particle. Very small bases
already give quite acceptable results, for the alpha particle, for example,
50 basis states give the binding energy within 1 MeV.  
The SVM results seem to be convergent 
in the model space 
defined with 300 basis states and  the partial-wave truncation with 
$\sum_{i=1}^{A-1} l_i \le 4$. 
We have tried to increase the accuracy by adding 700 more
states and by including  the  partial
waves up to $\sum_{i=1}^{A-1} l_i \le 6$ but the results 
are practically unchanged. The 1000 bases give only 2 keV 
gain in energy. The enlargement of the basis improves the 
expectation values, especially that of the kinetic energy operator, 
but this change is canceled by a similar change in the central 
potential. 
We think that the upper bound provided by the SVM calculation
is very close to the exact energy.
The accuracy 
achieved with few basis dimension is due to the use of the correlated 
Gaussian basis and the efficient optimization procedure.

\subsection{Hyperspherical Harmonic  Variational Method}

The Hyperspherical Harmonic (HH) functions constitute a general basis
for expanding the wave functions of an $A$-body
system~\cite{A89,F83,VKR95}. Very precise results can be obtained for
the three-nucleon bound state~\cite{KMRV96}.
In the HH variational method, the wave function is written as
\begin{equation}
  \Psi= \sum_\mu u_\mu(\rho) \Phi_\mu^{(K)}\ ,
\end{equation}
where $\rho$ is the hyper-radius. The quantities $\Phi_\mu^{(K)}$ 
are fully anti-symmetrized  HH-spin-isospin functions
of quantum numbers $\mu\equiv \{n,m,\ell_1,\ell_2,\ell_3,\ell_{12},L,
s,s',S,t,t',T\}$ constructed using the $K$-type Jacobi coordinates
${\bf x}_{1,2,3}$. Explicitly, 
\begin{eqnarray}
     \Phi^{(K)}_\mu&=& {\cal A}\Biggl[ (\sin\beta)^{2m}
           P_n^{\nu,\ell_{3}+{1\over2}}(\cos2\beta) 
           P_m^{\ell_{1}+{1\over2},\ell_{2}+{1\over2}}(\cos2\gamma)
          \;  \biggl [\Bigl[ \eta_t(12) 
              \eta(3)\Bigr]_{t'}
              \eta(4)  \biggr]_{TT_z} \times
          \nonumber \\
 \noalign{\medskip}
         && (x_{1})^{\ell_{1}} (x_{2})^{\ell_{2}} (x_{3})^{\ell_{3}}
          \biggl \{\Bigl [\bigl[ Y_{\ell_{1}}(\hat x_{1}) 
          Y_{\ell_{2}}(\hat x_{2})\bigr]_{\ell_{12}} 
            Y_{\ell_{3}}(\hat x_{3}) \Bigr ]_{L} 
          \; \biggl [\Bigl[\chi_s(12)
          \chi(3)\Bigr]_{s'} \chi(4) \biggr]_{S} \biggr
           \}_{JJ_z} \Biggr]\ , \label{eq:HH}
\end{eqnarray}
where $\cos\beta=x_3/\rho$, $\cos\gamma=x_2/(\rho\sin\beta)$ and
$\chi$ and $\eta$ denote spin and isospin functions, respectively.
Moreover, $\nu=\ell_1+\ell_2+2m+2$ and $P^{a,b}_n$  are Jacobi
polynomials (the integers $n$ and $m$ range from zero to infinity).
The coefficients $u_\mu(\rho)$  depend on the 
hyper-radius $\rho=\sqrt{(x_{1})^2+(x_{2})^2+(x_{3})^2}$
and can be  determined by solving a set of
second--order differential equations derived from the Rayleigh-Ritz
variational principle. For $A=4$ the necessary matrix elements of the
potential have been calculated by exploiting the techniques discussed
in Ref.~\cite{V98}.

The main difficulty in applying the HH technique to nuclear systems
is the very slow convergence of the expansion due to the strong
repulsion between the particles at short distances. In the $A=4$ case,
it has been found  convenient to separate
the HH states in different classes and to study the convergence by
including the states of one class at time~\cite{KRV01}.  The adopted
criterion has been to first include the HH functions
describing two-body correlations and, successively, those
incorporating three- and four-body correlations. Moreover, the
HH functions having the lowest orbital angular momentum quantum
numbers $\ell_i$ ($i=1,2,3$) have been included first.

\subsection{Green's Function Monte Carlo Method}

Green's function Monte Carlo methods use stochastic sampling 
to evaluate path integrals of the form:
\begin{eqnarray}
\Psi_0 &=& \lim_{\tau \rightarrow \infty} \Psi(\tau) \ , \\
\Psi(\tau) &=& e^{-(H-E_0)\tau} \Psi_T \ , \\
&=& \left[e^{-(H-E_0)\triangle\tau}\right]^{n} \Psi_T \ ,
\end{eqnarray}
where $\Psi_T$ is an approximate trial function obtained in a
variational or in an approximate constrained-path GFMC calculation
and  we have introduced a small time step, $\tau=n\triangle\tau$.
An approximate expression for the propagator, 
\begin{equation}
G({\bf R},{\bf R}^{\prime})= 
\langle {\bf R}|e^{-(H-E_0)\triangle\tau}|{\bf R}^{\prime}\rangle \ ,
\end{equation}
with error proportionate
to at least the second power of $\Delta\tau$ is used.
In the present work we use a symmetrized product of exact two-body
propagators, which has error proportionate to $(\Delta\tau)^3$
\cite{AV18+,aequal8}.

Green's function Monte Carlo calculations
for light nuclei with spin-isospin dependent interactions 
sample the particle coordinates while explicitly summing over 
the spin-isospin degrees of freedom~\cite{Carlson}.  The
first alpha particle calculations including $L \cdot S$ terms
employed the Reid V8 interaction~\cite{Carlson2}.  The chief
advantage of these methods is that they can be extended to
larger nuclei. More computationally efficient versions of the 
algorithm have been introduced and calculations extended up
to A=8~\cite{AV18+,aequal8}.

Convergence of the ground-state energy is governed by the
spectra of the Hamiltonian.  Calculations reported here
were performed to $\tau = 0.12\  {\rm MeV}^{-1}$.  
Since the first excited state of $^4$He is above 20 MeV,
any errors in $\Psi_T$ are damped out by at least 
$\exp(-2.4)$, an order of magnitude.  In fact our studies
show that the errors in  $\Psi_T$ correspond to much
higher excitation energies and $\langle H \rangle$ converges
by $\tau \sim 0.05{\rm MeV}^{-1}$.

The GFMC method allows us to compute mixed expectation
values of the form $\langle \Psi(\tau) | {\cal O} | \Psi_T \rangle$.
For $H$, this gives the exact ground state energy if $\tau$ is
large enough.
Expectation values of other quantities, such as pieces of
the Hamiltonian,
are often obtained through a linear extrapolation in the
error of the trial wave function:
\begin{equation}
\langle \Psi(\tau) | {\cal O} | \Psi(\tau) \rangle \approx
2 \langle \Psi(\tau) | {\cal O} | \Psi_T \rangle -
\langle \Psi_T | {\cal O} | \Psi_T \rangle,
\end{equation}
though it is possible to go beyond this approximation.

\subsection{No-Core Shell Model Method}

The NCSM is an approach applicable to both few-nucleon systems as well as
to light nuclei \cite{Navratil}. The calculations are performed in a finite
model space in the harmonic-oscillator (HO) basis. The model space ($P$) is spanned
by states with the total number of HO quanta $N\leq N_{\rm max}$. 
The Hamiltonian, 
\begin{equation}
H=T+V \ , 
\end{equation}
is modified by a HO center-of-mass
potential. Thus, we work with  
\begin{eqnarray}
H_A^\Omega &=& \sum_{i=1}^A \left[ \frac{p_i^2}{2m}
+\frac{1}{2}m\Omega^2 r^2_i
\right] \\
&+& \sum_{i<j=1}^A \left[ V(\vec{r}_i-\vec{r}_j)
-\frac{m\Omega^2}{2A}
(\vec{r}_i-\vec{r}_j)^2 
\right] \ .
\end{eqnarray}
As the NN potential depends on the relative coordinates, the added
HO term has no influence on the internal motion in the full space.
The effective Hamiltonian, appropriate to the finite $P$-space, 
is derived by the hermitian version of the Lee-Suzuki method \cite{LS80}.
In general, the effective Hamiltonian is an $A$-body operator.
We make an approximation 
by using just a two(three)-body effective interaction, which is obtained
by applying the Lee-Suzuki approach to the two(three)-nucleon system 
using $H_A^\Omega$ with the sums restricted to two(three) nucleons,
but with the $A$ in the interaction term kept fixed to, e.g., $A=4$ for $^4$He. 
Consequently, we deal with a two(three)-nucleon system bound in a HO potential.
The effective two(three)-body interaction, then replaces the interaction 
term in $H_A^\Omega$.
We note that the effective interaction, by construction, converges to the original
bare interaction as the basis space is increased and, thus, the NCSM calculation
converges to the exact solution with the basis-space enlargement. In fact,
it converges much faster than the corresponding bare interaction calculation
performed in the same basis.
Eventually, the $A$-nucleon $P$-space calculation can be performed
either in a Slater-determinant single-particle HO basis or in a 
properly antisymmetrized Jacobi-coordinate HO basis. The latter is used
in the present $^4$He calculation. In the past, we applied this approach successfully
to the $^4$He interacting by the CD-Bonn NN potential. It turns out that
the convergence with the AV8$'$ is significantly slower.
The limitation to a two-body effective interaction is
inadequate in the $P$-spaces that we could access ($N_{\rm max}=18$). 
Therefore, we performed the calculations using the three-body effective interaction.  
The mean values of the different operators were calculated using
the corresponding effective operators
computed within the Lee-Suzuki approach in a two-body approximation using 
the formula derived in Ref. \cite{LSeffop}. 
   
\subsection{Effective Interaction Hyperspherical Harmonic Method  }

Similarly to the preceeding method the EIHH approach introduces a two-body 
effective interaction $V_{eff}$ \cite{Barnea}. The division of the total HH 
space in P and Q spaces is realized via the HH quantum number $K$ (P(Q) space: 
$K\le(>)K_{max}$). Two powerful algorithms recently developed for the 
construction of symmetrized HH functions are employed \cite{BN1,BN2}.
In hyperspherical coordinates the total Hamiltonian is written as
\begin{equation}
  H =\frac{1}{2 m} \left(-\Delta_{\rho}+ \frac{\hat{K}^2}{\rho^2}\right) 
         + \sum_{i<j} V_{ij} \;,
\label{HH1}
\end{equation}
where $\rho$ is the hyper-radius and $\Delta_{\rho}$ contains derivatives with
respect to $\rho$ only. The grand-angular momentum operator $\hat{K}^2$ is a 
function of the variables of particles $A$ and $(A-1)$ and of $\hat K_{A-2}$ the 
grand angular momentum operator of the $(A-2)$ residual system \cite{Efros}.
Then from the total Hamiltonian one can extract a ``two-body'' Hamiltonian
of particles $A$ and $(A-1$)
\begin{equation}
  H_{2}(\rho) = \frac{1}{2 m} \frac{\hat{K}^2}{\rho^2} 
         + V_{A (A-1)} \;,
\label{HH2}
\end{equation}
which, however, contains the hyperspherical part of the total kinetic energy. 
Since the HH functions of the $(A-2)$ system are eigenfunctions of 
$\hat K^2_{A-2}$ one has an explicit dependence of $H_2$ on the quantum 
number $K_{A-2}$ of the residual system, i.e. $H_{2} \rightarrow 
H_{2}^{K_{A-2}}$.
Applying the hermitian version of the Lee-Suzuki method \cite{LS80}
to $H_2$ one gets an effective Hamiltonian $H_{2eff}$. The effective 
interaction $V_{eff}$ is obtained from 
\begin{equation}
 V_{eff}^{K_{A-2}}(\rho) = H_{2eff}^{K_{A-2}}(\rho) - \frac{1}{2 m} 
\frac{\hat{K}^2}{\rho^2}
\label{HH3}
\end{equation}
This $V_{eff}$ replaces $V_{ij}$ in Eq. (\ref{HH3}) when we project the solution
on the P-space.
This effective potential has the following property: $V_{eff}\to V_{ij}$ for 
$P\to 1$. Due to the ``effectiveness'' of the operator the solution of the 
Schr\"odinger equation converges faster to the true one. 
The HH formulation leads to various advantages: (i) $V_{eff}$ 
itself is $\rho$ dependent, therefore it contains some information on the 
``medium'', (ii) because of the above mentioned $K_{A-2}$ dependence
the (A-2) residual system is not a pure spectator, and (iii) an additional 
confining potential is not needed, since the presence of $\rho$ in Eq. 
(\ref{HH2}) automatically confines the two-body system to the range
$0\leq r_{A-(A-1)}< \sqrt{2} \rho$. We would like to point out that 
$V_{eff}(K_{max})$ can be viewed as a kind of momentum expansion, since the 
short range resolution is increased with growing $K_{max}$. As discussed
for the NCSM approach one obtains a better convergence for the calculation
of mean values introducing corresponding effective operators. 
Of course for the calculation of the mean value of the Hamiltonian, 
i. e. $E_b$, one already makes use of an effective operator, namely 
$H_{2eff}^{K_{A-2}}$.

\section{Results}

The AV8$'$ interaction 
appears to be an 
ideal test potential to compare the different calculational schemes. 
It is derived from the realistic AV18 interaction
\cite{AV18}  by neglecting the charge dependence and 
the terms proportional 
to $L^2$ and $(L\cdot S)^2$. Furthermore, in this work we omit the  electromagnetic
 part of the interaction. 
The potential is local and its spin and isospin dependences  are 
represented by operators. Because of its form it is tractable  for all 
of the calculational schemes described above. 

The potential consists of 8 parts:
\begin{eqnarray}
V(r) &=& V_c (r) +   V_\tau (r) 
(\tau \cdot \tau ) + V_\sigma (r) (\sigma \cdot \sigma ) +
V_{\sigma \tau}(r) (\sigma \cdot \sigma ) (\tau \cdot \tau )
\cr 
&&+ V_{t} (r) S_{12}  + V_{t \tau} (r) S_{12} \  (\tau \cdot \tau ) 
\cr
&& +
V_{b} (r) ( L\cdot S )
+ V_{b\tau}(r)   ( L\cdot S ) \  ( \tau \cdot \tau )
\label{eq16}
\\
&=& \sum _{i=1}^8 V_i (r) {\cal O}_i ,
\end{eqnarray}
where $(\sigma \cdot \sigma ) , (\tau \cdot \tau ), S_{12}$ and 
$ ( L\cdot S )$ 
stand for  
spin-spin, isospin-isospin, tensor and spin-orbit  interactions \cite{AV18}
 respectively, 
and $V_i (r)$ are radial functions of Yukawa- and Wood-Saxon types.
The AV18 and AV8' are defined with $\hbar ^2 /m_N$=41.47108MeV $fm^2$, computed from the average of the proton and neutron masses. Most of the results reported here were obtained using the traditional value of 41.47; this results in a change in $ \langle H  \rangle  $ of only $\approx$ 2.6 keV, far less than the estimated errors 
in the various methods.

First, we compare the binding energy results $E_b$,
the 
expectation values of the kinetic and potential energy and
 the radii in Table \ref{table2}.
We find good agreement for $E_b$ within 3 digits or within 0.5 \%. 
This is quite remarkable in view of the very  different techniques and the 
complexity of the nuclear force chosen. 
Except for  NCSM and EIHH, the expectation values of $T$ and $V $ also agree 
within 3 digits. The NCSM results are,  however,
 still within 1 \% and EIHH within 
1.5 \%of the others,
but note that the EIHH results for $T$ and $V$ are obtained with bare
operators.
The uncertainty in the NCSM results is of the same size, 
i.e., 1MeV, as that for the GFMC.
Finally, the given radii are also in very good agreement.

The HH calculation includes about 4500 states with
${\cal L}=\ell_1+\ell_2+\ell_3\le 6$.  The states with ${\cal L}=6$
give a contribution to the binding energy of approximately
$0.04$ MeV. 
It is to be noticed that the HH spin-isospin states $\Phi_\mu^{(H)}$
having ${\cal L}\le 6$ but constructed with the  $H$--type Jacobi
coordinates are linearly dependent on those  considered in
the expansion and therefore it is unnecessary to include
them. The contribution of $\Phi_\mu^{(K)}$ (and
$\Phi_\mu^{(H)}$) to the binding energy with ${\cal L}\ge 8$  has been
estimated to be approximately 0.01 MeV.

The errors quoted for the GFMC results are just the Monte Carlo statistical errors.
Various tests show that the energy is converged to at least this accuracy for
changes in $\Delta \tau$ or the maximum $\tau$.  There should be no other
sources of systematic error in this simple test case.

The NCSM binding energy result is based on extrapolation from calculations using
the three-body effective interaction in model spaces up to $N_{\rm max}=16$
in the HO frequency range $\hbar\Omega=16-43$ MeV. The mean values of different
operators, evaluated for $N_{\rm max}=16$ consisting of 2775 basis states and 
$\hbar\Omega=28$ MeV, 
were computed using effective operators as the
use of bare operators is completely insufficient, 
in particular for the $V_c(r)$ and $T$. Note that we have here 
$\langle T_{\rm eff}\rangle+\langle V_{\rm eff}\rangle$
close, but not exactly equal to $\langle H_{\rm eff}\rangle $,
due to approximations used. Overall, the NCSM results are less accurate
than the other methods. The NCSM  convergence rate is rather slow for the AV8'.
However, the method is flexible to  handle also non-local
realistic potentials like the CD-Bonn with a faster convergence 
rate due to a softer repulsive  core. The advantage of the method is
its applicability to the $p$-shell nuclei.

The EIHH calculation is carried out with $K_{max}=20$, (about 3000 HH states).
 The error estimate is based on the convergence with respect to $K_{max}$,
 i.e. difference of results for  $K_{max}=18$ and 20.
 An inspection of Table I 
shows that $E_b$ and radius are converged to a very high precision ($E_b$: 
0.04 \%, radius: 0.007 \%, not shown in Table I ). 
On the contrary $\langle T\rangle $ and $\langle V\rangle 
$ still change by about 
1 \% from $K_{max}=18$ to $K_{max}=20$.
Of course, by construction of the EIHH method, also  $\langle T\rangle $  
and $\langle V\rangle $
have to converge to the true result. In order to have a higher precision
one can proceed in two ways:
(i) increase of $K_{max}$, (ii) use of effective operators. Particularly
advantageous is the use of effective operators, since it allows us to make
rather precise calculations with a small number of basis functions
(see discussion of EIHH result for Fig. 1). As Table I shows it is not 
necessary to use
effective operators for long-range observables like the radius, while
observables that contain short range information (high momentum contributions),
like  $\langle T\rangle $ and $\langle V\rangle $, 
should, in principle, be calculated with effective operators.

A more detailed test of the wave function is to evaluate   
the expectation values of the eight  individual potential 
energy operators in eq. (\ref{eq16}).
The results are  shown in Table \ref{table3}. 
The agreement is, in general, rather good and well 
within 1 \%, except for NCSM with discrepancies up to 6 \% but they are 
generally  4\% or less.
In the case of the CRCGV, the expectation values for the spin-orbit operators 
are a bit off from the rest, but again still within 4 \%. There are no results
given for the EIHH. 

Table \ref{table4} shows the expectation 
values of the sum of the first 4 operators in eq. (\ref{eq16}) 
(called Central), of the two 
tensor operators and of the two spin-orbit operators. 
Again, no results are  given for the EIHH.
Except for the NCSM with differences up to 3.2 \%, 
all the values agree quite well each other.

As a further property of the wave function we consider the NN correlation 
function  
\begin{eqnarray}
C(r) = \langle  \Psi |  \delta ( \vec r - \vec r _{12} ) | \Psi \rangle ,
\end{eqnarray}
where 
$\vec r_{12}= \vec r_1 - \vec r_2 $. 
It is apparently normalized as 
 $ 4 \pi \int  C (r) r^2  dr = 1.$
The results for the various calculational schemes, except for the GFMC
 are shown in Fig. 1. 
The agreement among the FY, CRCGV, SVM, HH and NCSM is essentially perfect. 
For the EIHH it is necessary to use an effective operator in order to obtain 
good convergence also for 
$r<1.2$ fm. Due to the use of rather unsophisticated 
computers, the EIHH calculation for $C(r)$ is performed with the rather low 
$K_{max}$ value of 12 (about 400 HH states); 
however, a rather good agreement with the other methods is already obtained
at this low value.

Finally, we show in Table \ref{table5} the probabilities for 
finding the three different  total orbital angular momenta in our 4N  model 
system. 
The agreement among the different methods  is 
 very good with a small excursion in NCSM.

To summarize,  we have demonstrated that 
the Schr\"odinger equation for a four-nucleon system can be handled quite 
reliably by different methods leading to very good agreement in 
the binding energy, in expectation values of the 
kinetic and potential energies 
and in simple wave function properties. The AV8$'$ NN potential encompasses 
most of the complexity of realistic NN forces and, thus, the 
benchmark calculations are highly nontrivial and demonstrate the maturity 
and reliability of various methods. 
These results are 
 good foundations for further investigations 
of  nuclear structure for more complex systems 
and/or for other NN interaction models.

We have chosen the AV8$'$ potential because it can be handled without any approximation
by all of our methods. More realistic NN potentials such as AV18\cite{AV18}, 
CD-Bonn~\cite{cdBonn} and
Nijmegen I,II~\cite{Nijm}  pose additional difficulties for at least some of the methods.
There are new operator
forms with higher order derivatives or very strong nonlocalities. Also
some of the potentials are defined partial wave by partial wave.

Whereas in the four- body system the FY and NCSM schemes can handle
all types of NN potentials directly, the GFMC method relies on AV8' and
treats the difference to AV18 in perturbation theory. The SVM  can in
principle treat any local potential, such as AV18 , but the $(L\cdot S)^2$ terms require 
additional computational effort. 
Also the remaining methods, HH , CRCGV
and EIHH can handle more complicated potentials, 
although at present applications have been restricted to local potentials.
GFMC, NCSM, SVM and EIHH have already obtained
solutions for $A > 4$, whereas FY up to now has been restricted
to $A \leq 4$. An advantage of the  methods, CRCGV
and EIHH, is that they  do not need as heavy computational facilities as 
the other methods.

\begin{table} 
\caption{\label{table2}
The expectation values $\langle T  \rangle $ and $ \langle V  \rangle $ of kinetic and potential energies,
the binding energies $E_b$ in MeV 
and the radius in fm.
}
\begin{tabular}  {c|d|d|d|d} 
Method& { $\langle T  \rangle  $ } & {$\langle V  \rangle  $} & { $E_b$  } & { $\sqrt{ \langle r^2 \rangle  }$}\\
\hline
FY     & 102.39(5) & -128.33(10) & -25.94(5)   & 1.485(3) \\
CRCGV  & 102.30    & -128.20     & -25.90      & 1.482    \\
SVM    & 102.35    & -128.27     & -25.92      & 1.486    \\
HH     & 102.44    & -128.34     & -25.90(1)   & 1.483    \\
GFMC   & 102.3(1.0) & -128.25(1.0) & -25.93(2)  & 1.490(5) \\
NCSM   & 103.35     & -129.45     & -25.80(20) & 1.485    \\
EIHH   & 100.8(9)   & -126.7(9)  & -25.944(10) & 1.486   \\
\end{tabular}
\end{table}

\begin{table*}[tbp]
\caption{\label{table3}
Expectation values of the eight potential operators in Eq.(\protect\ref{eq16}) 
in MeV
}
\begin{tabular}  {c|d|d|d|d}   
Method & { $\langle V_c   \rangle $  } &  { $ \langle V_\tau  \rangle $  }  & { $\langle V_\sigma  \rangle$  } &  {  $\langle V_{\sigma \tau}  \rangle $  } \\  
\hline
FY     & 16.54& -5.038& -9.217& -57.55\\
CRCGV  & 16.54& -5.035& -9.215& -57.51\\
SVM    & 16.54& -5.036& -9.213& -57.51\\
HH     & 16.57& -5.034& -9.255& -57.59\\
GFMC   & 16.5(5)& -5.03(6) & -9.21(7)& -57.3(5) \\
NCSM   & 16.16& -4.92& -9.77& -57.89\\
\hline \hline
Method & { $ \langle V_{t}  \rangle $  } &  { $\langle V_{t \tau}  \rangle$  }  & { $ \langle V_{b} \rangle $  } &  { $\langle V_{b\tau}  \rangle $  } \\  
\hline
FY     & 0.707& -69.06& 10.79& -15.50\\
CRCGV  & 0.708& -68.99& 10.60& -15.30\\
SVM    & 0.707& -69.03& 10.78& -15.49\\
HH     & 0.702& -69.03& 10.76& -15.46\\
GFMC   &  0.71(3) & -68.8(5) &10.62(15) & -15.40(15) \\
NCSM   &  0.68& -69.13& 11.23& -15.80\\
\end{tabular}
\end{table*}

\begin{table} 
\caption{\label{table4}
Expectation values of potential energy operators in MeV.
}
\begin{tabular} {c|d|d|d}   
Method& { Central } & { Tensor  } &    { Spin-orbital  } \\
\hline
FY    &  -55.26~~~~~~&  -68.35~~~~~~&  -4.72~~~~~\\
CRCGV &  -55.22~~~~~~&  -68.28~~~~~~&  -4.70~~~~~~\\
SVM   &  -55.23~~~~~~&  -68.32~~~~~~&  -4.71~~~~~\\
HH    &  -55.31~~~~~~&  -68.32~~~~~~&  -4.71~~~~~\\
GFMC  &~~~~~-55.05(70) &~~~~-68.05(70) &~~~~-4.75(5)  \\
NCSM  &  -56.43~~~~~~&  -68.45~~~~~~&  -4.57~~~~~\\
\end{tabular}
\end{table}

\begin{table} 
\caption{\label{table5}
Probabilities of total orbital angular momentum components in [\%]
}
\begin{tabular}   {c|d|d|d}  
Method& { S-wave } & { P-wave } &    { D-wave } \\
\hline
FY    &  85.71~~~~~&  0.38~~~~~&  13.91~~~~~\\
CRCGV &  85.73~~~~~&  0.37~~~~~&  13.90~~~~~\\
SVM   &  85.72~~~~~&  0.368~~~~~&  13.91~~~~~\\
HH    &  85.72~~~~~&  0.369~~~~~&  13.91~~~~~\\
NCSM  &  86.73~~~~~&  0.29~~~~~~&  12.98~~~~~\\
EIHH  &  85.73(2) &  0.370(1) &  13.89(1) \\
\end{tabular}
\end{table}

\begin{figure}[htbp]
\begin{center}
\caption{Correlation functions in the different calculational schemes:
EIHH(dashed-dotted),  FY, CRCGV, SVM,  HH and NCSM (overlapping curves)
}
\mbox{\epsfxsize=90mm \epsffile{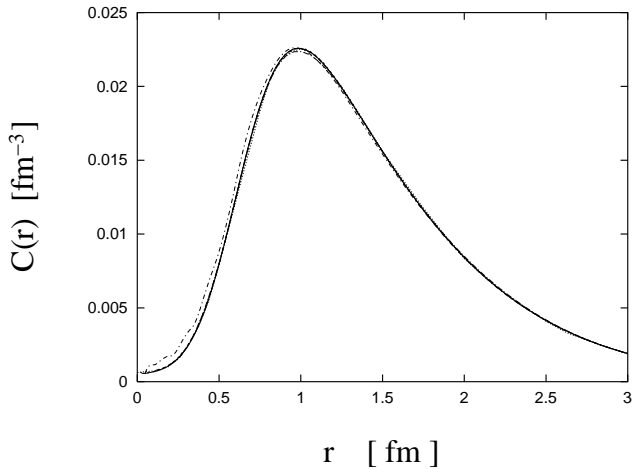}} 
\label{fig:1}
\end{center}
\vspace{-0.5cm}
\end{figure}

\bigskip

\acknowledgements
The work of A. N. was supported by the Deutsche Forschungsgemeinschaft. 
The numerical FY 
calculations have been performed on the CRAY T90 and  T3E of the 
John von Neumann Institute for Computing 
in J\"ulich, Germany.
The work of J. C. is supported by the U. S. Department of Energy under contract W-7405-ENG-36 and that of S.C.P. and R.B.W. by the U.S. Department of Energy, Nuclear Physics Division, under contract No. W-31-109-ENG-38. 
The GFMC calculations were made on the parallel computers of the Argonne 
Mathematics and Computer Science Division. 
  The work of P.N. was done under auspices of the U. S. Department of Energy
by the Lawrence Livermore National Laboratory under contract No. W-7405-ENG-48
and supported from LDRD contract 00-ERD-028.
B.R.B. acknowledges the NSF grant No. PHY-0070858.
The work of K.V. and Y.S was supported by  the
JSPS-HAS collaboration (2000-2002), Yamada Science Foundation
and a OTKA grant of No. T029003.
The work of W.L. and G.O. was supported by the Italian Ministry for 
Scientific and Technological Research (MURST).

\end{document}